\documentclass[aps,twocolumn]{revtex4}%
\usepackage{amsfonts}
\usepackage{amsmath}
\usepackage{amssymb}
\usepackage{graphicx}%
\begin{document}
\title{Ground state and excitation properties of the quantum kagom\'{e} system
ZnCu$_{3}$(OH)$_{6}$Cl$_{2}$ investigated by local probes.}
\author{Oren Ofer and Amit Keren}
\affiliation{Physics Department, Technion, Israel Institute of Technology, Haifa 32000, Israel}
\author{Emily A. Nytko, Matthew P. Shores, Bart M. Bartlett, and Daniel G. Nocera}
\affiliation{Department of Chemistry, Massachusetts Institute of Technology, Cambridge, MA
02139 USA}
\author{Chris Baines and Alex Amato}
\affiliation{Paul Scherrer Institute, CH 5232 Villigen PSI, Switzerland}

\begin{abstract}
We characterize the ground state and excitation spectrum of the $S=1/2$,
analytically pure and perfect kagom\'{e} system ZnCu$_{3}$(OH)$_{6}$Cl$_{2}$
using the following measurements: magnetization, muon spin rotation frequency
shift $K$, transverse relaxation time $T_{2}^{\ast}$, and Cl nuclear
spin-lattice relaxation $T_{1}$. We found no sign of singlet formation, no
long range order or spin freezing, and no sign of spin-Peierls transition even
at temperatures as low as $60$~mK. The density of states has an $E^{1/4}$
energy dependence with a negligible gap to excitation.

\end{abstract}
\maketitle

The study of spin $1/2$ quantum magnetism on the kagom\'{e} lattice is very
intriguing and lively because different investigation strategies have led to
fundamentally different predictions regarding the ground state and excitations
of this system. Theories based on numerical or approximate diagonalization of
the Heisenberg Hamiltonian favors non-magnetic (singlet) ground state
\cite{LecheminantPRB97,BudnikPRL04} with full symmetry of the Hamiltonian
\cite{ChalkerPRB92}, negligible correlation length \cite{LeungPRB 93}, and an
upper limit on excitation gap of $J/20$ \cite{WaldtmannEurPhyJ98}. These
strategies provide conflicting reports on the possibility of a spin-Peierls
state \cite{ChalkerPRB92,MarstonJAP91}. In contrast, semiclassical treatments
based on the thermal \cite{ChalkerPRL92} or quantum \cite{ChubukovPRL92}
\textquotedblleft order from disorder\textquotedblright\ approach gives rise
to a ground state with a broken symmetry of the $\sqrt{3}\times\sqrt{3}$ type,
which is stable even against tunneling \cite{DelftPRB93}, with gapless magnon
excitations \cite{HarrisPRB92}, and no spin-Peierls type distortion
\cite{KagomeDistortion}. In these circumstances, one expects experiments to
provide some guidance. However, the experimental situation is equally
confusing since most of the experimental studies of kagom\'{e}-like materials
suffered from several kinds of shortcomings, which made comparison with
theoretical models difficult. Some material, such as the jarosites
\cite{amitJaro} have spin $S>1/2$. Others have $S=1/2$ but with a non perfect
kagom\'{e} structure, such as the volborthite Cu$_{3}$V$_{2}$O$_{7}$(OH$_{2}%
$)2H$_{2}$O (CVO) \cite{fukaya}. A third class of materials such as
SrCr$_{9p}$Ga$_{12-9p}$O$_{19}$ \cite{amitSCGO1} and Ba$_{2}$Sn$_{2}%
$ZnGa$_{10-7p}$Cr$_{7p}$O$_{22}$ \cite{MendelsSCGO} were hampered by disorder
or strong third direction interaction. In light of these difficulties, the
recent synthesis \cite{shores1} of herbertsmithite, ZnCu$_{3}$(OH)$_{6}%
$Cl$_{2}$, with its Cu-based quantum $S=1/2$, analytically pure and perfect
kagom\'{e} lattice, should put the field on a new course.

In this letter we present a comprehensive study of ZnCu$_{3}$(OH)$_{6}$%
Cl$_{2}$ using local probes. In our study, we address four questions which are
at the heart of the investigation of the quantum kagom\'{e} system: Do $S=1/2$
spins on kagom\'{e} lattice freeze? Is the ground state magnetic? What is the
density of excited states, and is there a gap in the spin energy spectra?
Finally, does the lattice distort in order to accommodate spin-Peierls state?
We address these questions in the present work using nuclear magnetic
resonance (NMR) and muon spin resonance ($\mu$SR) local probes. We also use
magnetization measurements to calibrate the local probes.

ZnCu$_{3}$(OH)$_{6}$Cl$_{2}$ was prepared by hydrothermal methods performed at
autogenous pressure. A 800 mL teflon liner was charged with 16.7~g Cu$_{2}%
$(OH)$_{2}$CO$_{3}$ (75.5 mmol), 12.2~g of ZnCl$_{2}$ (89.5 mmol), and 350 mL
water, capped and placed into a custom-built steel hydrothermal bomb under
ambient room atmosphere. The tightened bomb was heated at a rate of
$1$~$^{\circ}$C/min to $210$~$^{\circ}$C and the temperature was maintained
for 48~h. The oven was cooled to room temperature at a rate of $0.1$~$^{\circ
}$C/min. A light blue powder was isolated from the base of the liner by
filtration, washed with water, and dried in air to afford 21.0~g product (49.0
mmol, 97.4\% yield based on starting Cu$_{2}$(OH)$_{2}$CO$_{3}$). Magnetic and
pXRD data were consistent with those previously reported for herbertsmithite
\cite{shores1}.

\begin{figure}[ptb]
\begin{center}
\includegraphics[width=8cm]{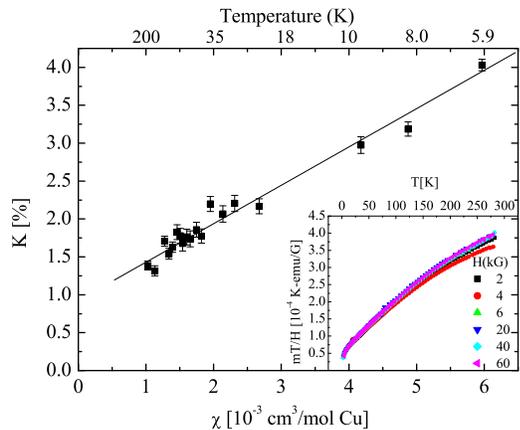}
\end{center}
\caption{The muon shift $K$ against susceptibility. In the inset, normalized
magnetization versus temperature.}%
\label{suscep}%
\end{figure}

DC magnetization $m$ measurements were performed on a powdered sample using a
Cryogenic SQUID magnetometer at temperatures $T$ ranging from $2$ to $280~$K
and fields $H$ varying from $2~$kG to $60~$kG. In the inset of Fig.
\ref{suscep} we present $mT/H$ versus $T$. The data collapse onto a single
line, especially at low $T$, meaning that the susceptibility is
field-independent in our range of temperatures and fields. Also, no peak in
the susceptibility is observed, indicating the absence of magnetic ordering.
The only indication of interactions between spin in these measurements is the
fact that $mT/H$ decreases upon cooling whereas in an ideal paramagnetic
system this quantity should be constant. Fits of the susceptibility data at
high temperatures and lower fields than presented here reveal a Curie-Weiss
temperature of $\Theta_{CW}=-314$~K \cite{HeltonPrivCom}. The frustration
parameter $T_{F}/|\Theta_{CW}|\approx0.22$, where $T_{F}$ $\sim70$~K is the
temperature at which $\chi^{-1}$ is no longer a linear function of $T$,
indicates strong geometric frustration. For comparison, in CVO $T_{F}%
/|\Theta_{CW}|\approx1$ \cite{hiroi}.

Muon spin rotation and relaxation ($\mu$SR) measurements were performed at the
Paul Scherrer Institute, Switzerland (PSI) in the GPS spectrometer with an He
cryostat, and in the LTF spectrometer with a dilution refrigerator. The
measurements were carried out with the muon spin tilted at $45^{\circ}$
relative to the direction of the beam. Positron data were collected in both
the forward-backward (longitudinal) and the up-down and right when available
(transverse) detectors simultaneously. Data were collected at temperatures
ranging from 60~mK to 200~K with a constant field of 2~kG. In GPS we used a
resistive magnet with a calibrated field; in LTF we used a superconducting
magnet. We took data at overlapping temperatures to calibrate the LTF field.
The longitudinal data, which measure muon spin $T_{1}$, were found to be
useless since the $T_{1}$ in our field range is much longer than the muon
lifetime and showed no temperature variations.

\begin{figure}[ptb]
\begin{center}
\includegraphics[width=8cm]{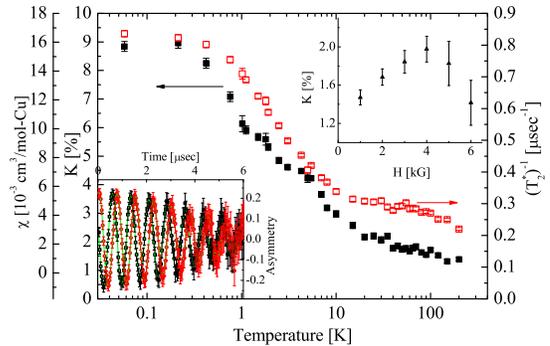}
\end{center}
\caption{A plot of the muon shift $K$, transverse relaxation time $\sigma$,
versus temperature. In the upper inset, the muon shift $K$ versus external
field $H$. In the lower inset, real and imaginary transverse field asymmetry
for $T=100$~K.}%
\label{musr}%
\end{figure}

In the lower inset of Fig. \ref{musr} we show real and imaginary transverse
field [TF] data taken at $H=2~$kG and $T=100$~K. The data are presented in a
rotating reference frame (RRF) at a field of $1.9~$kG. The TF asymmetry is
best described by $A_{TF}=A_{0}\exp\left(  -t^{2}/(2T_{2}^{\ast2})\right)
\cos(\omega t+\phi)$ where $T_{2}^{\ast}$ is the transverse relaxation time,
and $\omega$ is the frequency of the muon. The quality of the fit is
represented by the solid line.

In Fig.~\ref{suscep} we depict the frequency shift, $K=(\omega_{0}%
-\omega)/\omega_{0}$ where $\omega_{0}$ is the free muon rotation frequency in
the RRF. The difference in frequency between free and implanted muons is a
consequence of the sample magnetization; therefore, the shift is expected to
be proportional to the susceptibility. Indeed, as shown in the main panel of
Fig.~\ref{suscep}, there is a linear relation between $K$ and the
susceptibility $\chi=m/H$, with the temperature as an implicit parameter; some
representative temperatures are shown on the upper axis. In the upper inset of
Fig.~\ref{musr} we present the field dependence of the shift at $T=10$~K.
Within the error bars, the shift in the RRF does not depend on the external
field. This is in agreement with the susceptibility results. Therefore, Fig.
\ref{musr} could serve as a conversion graph from muon spin frequency shift to susceptibility.

In the main panel of Fig.~\ref{musr}, we depict $K$ as a function of
temperature. An additional axis is presented where $K$ has been converted to
$\chi$ as discussed above. We find that $K$ (and hence $\chi$) increases with
decreasing temperatures and saturates below $T\sim200~$mK at a value of
$\chi=15.7(5)\times10^{-3}$ cm$^{3}$/mol Cu; the error is from the
calibration. It should also be pointed out that the energy scale associated
with spin $1/2$ in a field of $2$~kG is $200~$mK, and the saturation could be
a consequence of the external field. The saturation of $\chi$ is a strong
evidence for the lack of impurities in our sample. More importantly, it
indicates the lack of singlet formation or spin freezing. The last conclusion
is also in agreement with neutron scattering measurements \cite{HeltonPrivCom}
and zero field $\mu$SR \cite{mendelscondmat}.

The muon transverse relaxation rate $1/T_{2}^{\ast}$ is also presented in Fig.
\ref{musr}. Roughly speaking, it has the same temperature behavior as the
shift (and as the susceptibility). $T_{2}^{\ast}$ relaxation is a result of
defects in the sample causing a distribution of muons to electronic spin
coupling constants or a distribution of susceptibilities. It has been shown
that when the muon relaxation rate behaves similarly to the shift \cite{tbtio}
(or susceptibility \cite{eva}) upon cooling, it indicates quenched
distribution of either the coupling constants or susceptibilities. In this
case the relaxation increases simply because the average moment size
increases. Since the coupling constants and susceptibility are functions of
distances between muon and electronic spin or between two electronic spins,
our results are consistent with a lack of lattice deformation in ZnCu$_{3}%
$(OH)$_{6}$Cl$_{2}$.

\begin{figure}[tb]
\begin{center}
\includegraphics[width=8cm]{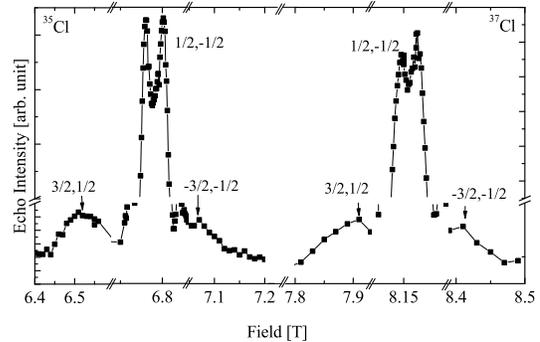}
\end{center}
\caption{A field sweep of $^{35}$Cl and $^{37}$Cl.}%
\label{nmrspec}%
\end{figure}

Since muons could not reveal dynamic $T_{1}$ information, we performed $^{37}%
$Cl and $^{35}$Cl NMR experiments on the same sample. Using the two isotopes,
we are able to determine the origin of $T_{1}$. The first step in such a
measurement is to find the line shape and to identify the isotopes and
transitions. This measurement was done at a constant applied frequency of
$\nu_{app}=28.28$~MHz and a varying external field $H$. A standard spin-echo
pulse sequence, $\pi/2-\tau-\pi$, was applied, and the echo signal was
integrated for each $H$. In Fig. \ref{nmrspec} we show a field sweep for both
Cl isotopes obtained at $T=100$~K. A rich spectrum is found and is emphasized
using five $x$-axis and one $y$-axis breakers. This rich spectrum is a
consequence of the Cl having spin $3/2$ for both isotopes. In the case where
the nuclei reside in a site with non cubic local environment and experience an
electric field gradient, their spin Hamiltonian could be written as
$\mathcal{H}=-h\nu_{l}\mathbf{I}\cdot(1+\mathbf{K})\cdot\widehat{\mathbf{H}%
}+(h\nu_{Q})/6\left[  3I_{z}^{2}-I^{2}+\eta\left(  I_{x}^{2}-I_{y}^{2}\right)
\right]  $ where $\nu_{Q}$ is the quadrupole frequency, $0\leq\eta\leq1$ is
the anisotropy parameter, $\mathbf{K}$ is the shift tensor, and $\nu
_{l}=\gamma H/(2\pi)$. The powder spectrum of such nuclei has two satellite
peaks corresponding to the $3/2\longleftrightarrow1/2$ and
$-3/2\longleftrightarrow-1/2$ transitions, and a central line from the
$1/2\longleftrightarrow-1/2$ transition, which is split due to the powder
average. The transition names are presented in the figure. The satellite peaks
at $T=100$~K of $^{35}$Cl are at $6.52$ and $7.07$~T, and for $^{37}$Cl at
$7.91$ and $8.41$ T. The lack of singularity in the satellite spectrum
indicates that the Cl resides in a site with $\eta>0$, namely, with no $xy$
symmetry. These findings are consistent with a 3m site symmetry of the Cl ion
in the space group R$\bar{3}$m.

In contrast to the two satellites, the spliting of the central lines at
$T=100$~K is clear, and appear for the $^{35}$Cl at $6.778$ and $6.801$~T and
for the $^{37}$Cl at $8.148$ and $8.161$~T. Under some assumptions these
values could be used to determine the parameter of the nuclear spin
Hamiltonian \cite{amitSCGO1}; assuming that the nuclear spin operators,
$I_{x},$ $I_{y}$ and $I_{z}$ are colinear with the principal axes of the shift
tensor, and that the in-plane shift is isotropic with $K_{\perp}=(K_{x}%
+K_{y})/2$, we find for both isotopes, $K_{\perp}\simeq-0.0017(5)$,
$K_{z}\simeq0.035(9)$ and $\eta=0.4$, and $^{35}\nu_{Q}=3.75$~MHz and
$^{37}\nu_{Q}=2.55$~MHz. The ratio of $\nu_{Q}$ is as expected from the ratio
of the quadrupole moments. Due to the assumptions, the value of $K_{z}$ should
only be considered as an order of magnitude. Nevertheless, it is not very
different from that of the muon shift (in the laboratory frame) at the same
temperatures. This means that both probes experience a similar field generated
by the Cu spins, which for muons is usually a dipolar field.

\begin{figure}[b]
\begin{center}
\includegraphics[width=8cm]{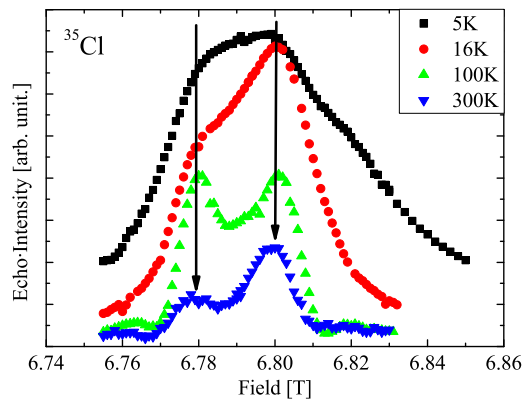}
\end{center}
\caption{$^{35}$Cl field sweep ($\nu=28.28MHz$) at different temperatures. The
arrows indicate the central line singularities observed at high-$T$ but
smeared out at low $T$.}%
\label{nmrshift}%
\end{figure}

Temperature dependence field sweeps of the $^{35}$Cl central line are shown in
Fig. \ref{nmrshift}. The intensities are in arbitrary units for clarity. The
$\pm1/2\longleftrightarrow\mp1/2$ transitions are easily observed at $T=300$
and $100~$K (indicated by the arrows in the figure) but are smeared out at
lower $T$. In fact, the lines become so broad that the NMR shift cannot be
followed to low temperature; hence the importance of the $\mu$SR results.

\begin{figure}[ptb]
\begin{center}
\includegraphics[width=8cm]{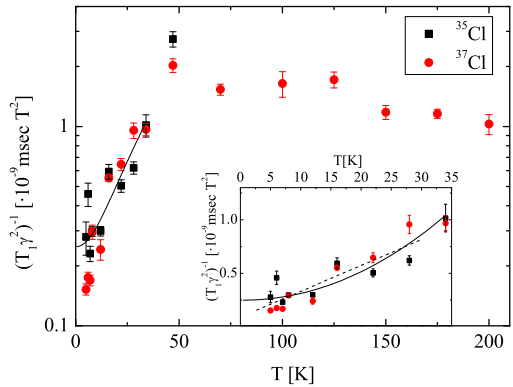}
\end{center}
\caption{A semi-log plot of the Cl inverse spin-lattice relaxation,
$(\gamma^{2}T_{1})^{-1}$, versus temperature. Inset, a linear plot of the
low-temperature region. The black line is a fit to Eq.~\ref{1overt1}. The
dashed line is straight.}%
\label{nmrt1}%
\end{figure}

Finally, we measured the $^{37}$Cl spin-lattice relaxation rate $T_{1}^{-1}$
to determine spin gap and excitation spectrum. The data were taken at a field
of 8.15~T which corresponds to the low field peak of the central line. We use
a saturation recovery pulse sequence. In Fig. \ref{nmrt1} we depict
$T_{1}^{-1}$ normalized by $\gamma^{2}$ where $^{37}\gamma=3.476$ MHz/T on a
semi-log scale. $T_{1}^{-1}$ increases upon cooling down to $50$~K and then
sharply decreases. We also present $^{35}$Cl $(T_{1}\gamma^{2})^{-1}$ where
$^{35}\gamma=4.172$ MHz/T below $50$~K in order to determine the origin of the
dynamic fluctuations. These measurements were done under the same conditions
as $^{37}$Cl. When considering all temperatures we find that $T_{1}^{35}%
/T_{1}^{37}=0.75(10)$. From a magnetic relaxation mechanism we expect this
ratio to equal $(^{37}\gamma/^{35}\gamma)^{2}=0.69$. From a quadrupole based
mechanism we anticipate $(^{37}Q/^{35}Q)^{2}=0.62$ where $Q$ is the nuclear
quadrupole moment. Our finding is in favor of relaxation mediated by a
magnetic mechanism as indicated by the overlapping $(T_{1}\gamma^{2})^{-1}$
data points in Fig. \ref{nmrt1}. 

In the inset of Fig. \ref{nmrt1} we zoom in on the low $T$ data using a linear
scale. A first glance suggests that at low temperature $1/T_{1}$ is a linear
function of $T$ as indicated by the dashed line. A remanent relaxation at zero
temperature $1/T_{1}^{n}$ could be due to magnetic fluctuations form other
nuclear moments such as the protons or copper, since they continue to
fluctuate even when the electronic moments stop. Thus the spin lattice
relaxation due to electronic contribution only $1/T_{1}^{e}\equiv
1/T_{1}(T)-1/T_{1}^{n}$ obeys $1/(T_{1}^{e}T)=const$. This relation is
expected in the case of free fermions and might be related to recent theories
\cite{Fermions,SpinLiquieds}.

A different approach to $T_{1}$ interpretation is in terms of magnon Raman
scattering where
\begin{equation}
\frac{1}{T_{1}}(T)=\frac{1}{T_{1}^{n}}+\gamma^{2}A^{2}\int_{\Delta}^{\infty
}\rho^{2}(E)\cdot n(E)\cdot\left[  n(E)+1\right]  dE \label{1overt1}%
\end{equation}
with $\rho$ being the density of states, $\Delta$ the gap, $A$ is
a constant derived from the hyperfine coupling, and $n(E)$ the
Bose-Einstein occupation factor \cite{KranendonkPhys56}. This
expression is constructed from the population of magnons before
and after the scattering, with the associated density of states
and the assumption that they exchanged negligible amount of energy
with the nuclei since its Zeeman splitting is much less than a
typical magnon energy. However, in frustrated magnets the magnon
might not be the proper description of the excitations
\cite{amitSCGO1,MendelsSCGO}. Nevertheless, we use
Eq.~\ref{1overt1} since it is expected for any kind of bosonic
excitations, and since there is no other available theory. We
assume $\rho(E)\sim E^{\alpha}$, with $\alpha$ and $\Delta$ as fit
parameters. The fit of Eq.~\ref{1overt1} to the data is presented
as the solid line in Fig.~\ref{nmrt1}, and in its inset. We find
$\alpha=0.23(1)$ and $\Delta=0.5(2)$ K. Comparing to $J=209$~K
\cite{HeltonPrivCom}, this is a negligibly small gap. It indicates
that most likely there is no gap in the spin energy spectra, in
agreement with Ref.~\cite{HeltonPrivCom}, and $\rho(E)\sim
E^{1/4}$.

To conclude, susceptibility measurements down to 60~mK
suggest that there is no freezing and only a saturation of
susceptibility, namely, no singlet formation. The data also do not
support the presence of lattice deformation. Finally, Cl NMR
$T_{1}$ measurements find a negligibly small magnetic gap and the
density of states $\rho\sim E^{1/4}$. Thus,
ZnCu$_{3}$(OH)$_{6}$Cl$_{2}$ is an exotic magnet with no broken
continuous symmetry but gapless excitations. It might be an
example of algebraic spin liquid \cite{SpinLiquieds}.

We would like to thank the PSI facility for supporting the $\mu$SR
experiments and for continuous high quality beam, and the NATO
Collaborative Linkage Grant, reference number PST.CLG.978705. We
acknowledge helpful discussions with Young. S. Lee and Philippe.
Mendels, Peter Mueller, Joel Helton, and Kittiwit Matan.


\begin{thebibliography}{99}                                                                                               %


\bibitem {LecheminantPRB97}P. Lecheminant, B. Bernu and C. Lhuillier, L.
Pierre, and P. Sindzingre, Phys. Rev. B \textbf{56}, 2521 (1997).

\bibitem {BudnikPRL04}R. Budnik and A. Auerbach, Phys. Rev. Lett. \textbf{93},
187205 (2004).

\bibitem {ChalkerPRB92}J. T. Chalker and J. F. G. Eastmond, Phys. Rev. B
\textbf{46}, 14201 (1992).

\bibitem {LeungPRB 93}P. W. Leung and V. Elser, Phys. Rev. B \textbf{47}, 5459 (1993).

\bibitem {WaldtmannEurPhyJ98}Ch. Waldtmann, H.-U. Everts, B. Bernu, C.
Lhuillier, P. Sindzingre, P. Lecheminant, L. Pierre, Eur. Phys. J. B
\textbf{2}, 501 (1998).

\bibitem {MarstonJAP91}J. B. Marston and C. Zeng, J. Appl. Phys, \textbf{69}
5962 (1991).

\bibitem {ChalkerPRL92}J. T. Chalker, P. C. W. Holdsworth, and E. F. Shender,
Phys. Rev. Lett. \textbf{68} 855 (1992).

\bibitem {ChubukovPRL92}A. Chubukov, Phys. Rev. Lett. \textbf{69}, 832 (1992).

\bibitem {HarrisPRB92}A. B. Harris, C. Kallin, and A. J. Berlinsky Phys. Rev.
B \textbf{45}, 2899 (1992).

\bibitem {DelftPRB93}J. von Delft and C. L. Henley, Phys. Rev. B \textbf{48},
965 (1993).

\bibitem {KagomeDistortion}C. Jia, J. H. Nam, J. S. Kim, and J. H. Han, Phys.
Rev. B \textbf{71}, 212406 (2005); O.~Tchernyshyov, private communication.

\bibitem {amitJaro}A. Keren, K. Kojima, L. P. Le, G. M. Luke, W. D. Wu, and Y.
J. Uemura, M. Takano, H. Dabkowska, M. J. P. Gingras, Phys. Rev. B \textbf{53}
6451 (1996).

\bibitem {amitSCGO1}A. Keren Y. J. Uemura, G. Luke, P. Mendels, M. Mekata, and
T. Asano, Phys. Rev. Lett. \textbf{84}, 3450 (2000).

\bibitem {MendelsSCGO}D. BonoD. Bono, P. Mendels, G. Collin, N. Blanchard, F.
Bert, A. Amato, C. Baines, and A. D. Hillier, Phys. Rev. Lett. \textbf{93},
187201 (2004).

\bibitem {fukaya}A. Fukaya, Y. Fudamoto, I. M. Gat, T. Ito, M. I. Larkin, A.
T. Savici, Y. J. Uemura, P. P. Kyriakou, G. M. Luke, M. T. Rovers, K. M.
Kojima, A. Keren, M. Hanawa, and Z. Hiroi, Phys. Rev. Lett. \textbf{91} 207603 (2003).

\bibitem {shores1}Matthew P. Shores, Emily A. Nytko, Bart M. Bartlett and
Daniel G. Nocera, J. Am. Chem. Soc., \textbf{127 }, 13462 (2005).

\bibitem {HeltonPrivCom}J.S. Helton, K. Matan, M.P. Shores, E.A. Nytko, B.M.
Bartlett, Y. Yoshida, Y. Takano, Y. Qiu, J.-H. Chung, D.G. Nocera, Y.S. Lee, cond-mat/0610539.

\bibitem {hiroi}Z. Hiroi, M. Hanawa, N. Kobayashi, M. Nohara, H. Takagi, Y.
Kato and M. Takigawa, J. Phys. Soc. Japan \textbf{70}, 3377 (2001).

\bibitem {mendelscondmat}P. Mendels, F. Bert, M.A. de Vries, A. Olariu, A.
Harrison, F.Duc, J.C. Trombe, J. Lord, A. Amato and C. Baines, cond-mat/0610565.

\bibitem {tbtio}Oren Ofer, Amit Keren, Chris Baines and Jason S Gardner, To be
published in J. Phys.: Condens. Matter.

\bibitem {eva}Eva Sagi, Oren Ofer and Amit Keren, Jason S Gardner, Phys. Rev.
Lett. \textbf{94}, 237202 (2005).

\bibitem {Fermions}O. I. Motrunich, Phys. Rev. B \textbf{72}, 045105 (2005);
S. S. Lee and P. A. Lee, Phys. Rev. Lett. \textbf{95}, 036403 (2005).

\bibitem {SpinLiquieds}Y. Zhou and X. Wen, cond-mat/02106662; J. Alicea, O. I.
Motrunich, and M. P. A. Fisher, Phys. Rev. Lett, \textbf{95}, 247203 (2005).

\bibitem {KranendonkPhys56}J. Van Kranendonk and M. Bloom Physica
\textbf{XXII}, 545 (1956).
\end{thebibliography}
\end{document}